\documentclass[12pt,a4,onecolumn,secnumarabic,amssymb, amsmath, nofootinbib,tightenlines,
nobibnotes, aps, prl,epsfig]{revtex4}
\usepackage{graphicx}
\usepackage{dcolumn}
\usepackage{bm}
\begin{document}
\preprint{APS/123-QED}
\title{ Hard- Pomeron behavior of the Longitudinal Structure Function $F_{L}$ in the Next- to- Leading- Order at low $x$}

\author{G.R.Boroun }
\altaffiliation{boroun@razi.ac.ir}
\author{}%
\affiliation{ Physics Department, Razi University, Kermanshah
67149, Iran}
\date{\today}

\begin{abstract}
We present an analytic formula to extract the longitudinal
structure function in the next- to -leading order of the
perturbation theory at low $x$, from the Regge- like behavior of
the gluon distribution and the structure function at this limit.
In this approach, the longitudinal structure function has the
hard- Pomeron behavior. The determined values are compared with
the $H1$ data and MRST model. All results can consistently be
described within the framework of perturbative QCD which
essentially show increases as $x$ decreases.
\end{abstract}
\maketitle
\subsection{1 Introduction}
The longitudinal structure function $F_{L}(x,Q^{2})$ comes as a
consequence of the violation of Callan- Gross relation [1] and is
defined as $F_{L}(x,Q^{2})=F_{2}(x,Q^{2})-2xF_{1}(x,Q^{2})$, Where
$F_{2}(x,Q^{2})$ is the transverse structure function. As usual
$x$ is the Bjorken scaling parameter and $Q^{2}$ is the four
momentum transfer in a deep inelastic scattering process. In the
quark parton model (QPM) the structure function $F_{2}$ can be
expressed as a sum of the quark- antiquark momentum distributions
$xq_{i}(x)$ weighted with the square of the quark electric charges
$e_{i}$:
$F_{2}(x)={\sum_{i}}e_{i}^{2}x(q_{i}(x)+\overline{q}_{i}(x))$. For
spin $\frac{1}{2}$ partons QPM also predicts $F_{L}(x)=0$,
which leads to the Callan- Gross relation.\\

The naive QPM has to be modified in QCD as quarks interact through
gluons, and can radiate gluons. Radiated gluons, in turn, can
split into quark- antiquark pairs (sea quarks) or gluons. The
gluon radiation results in a transverse momentum component of the
quarks. Thus, in QCD the longitudinal structure function is non-
zero. Due to its origin, $F_{L}$ is directly dependent to the
gluon distribution in the proton and therefore the measurement of
$F_{L}$ provides a sensitive test of perturbative QCD [2]. The
next- to- leading order (NLO) corrections to the longitudinal
structure function are large and negative, valid to be at small
$x$ [3-5].\\

At small $x$, the longitudinal structure function can be related
to the gluon and sea- quark distribution.
 In principle, the data on the singlet part
 of the structure function $F_{2}$ constrain the sea quarks
  and the data on the slope $\frac{dF_{2}}{d{\ln}Q^{2}}$
  determine the gluon density [6-10]. One of the most striking discoveries at HERA is the steep rise of
the proton structure function $F_{2}(x,Q^{2})$ with decreasing
Bjorken $x$ [11]. The behavior of the structure function at small
$x$ is driven by the gluon through the process
$g\hspace{0.1cm}{\rightarrow}\hspace{0.1cm}q\overline{q}$.
Therefore the gluon distribution is observed that governs the
physics of high energy processes in QCD. HERA shows that the steep
inelastic structure function $F_{2}(x,Q^{2})$ has a steep behavior
in the small x region ($10^{-2}{>}x{>}10^{-5}$), even for very
small virtualities ($Q^{2}{\approx}1 GeV^{2}$). This steep
behavior is well described in the framework of the DGLAP [12]
evolution equations. So, we restrict our investigations to the
Regge- like behavior for the gluon distribution and the structure
function by the following forms:
\begin{equation}
F_{2}(x,Q^{2})=\hspace{0.1cm}xS \hspace{0.1cm}{\sim}\hspace{0.1cm}
A_{S}x^{-\lambda_{S}},
\end{equation}
and
\begin{equation}
xg(x,Q^{2})=\hspace{0.1cm} A_{g}x^{-\lambda_{g}}.
\end{equation}\\

The singlet part of the structure function is controlled by
Pomeron exchange at small $x$, where $\lambda_{S}$ is the Pomeron
intercept minus one. The exponent was rapid rise in $Q^{2}$ of the
structure function at this limit. This steep behavior of the
structure function generates a similar steep behavior of the gluon
distribution at small $x$, where $\lambda_{S}{\neq}\lambda_{g}$ in
next- to- leading order analysis. We also note that $\lambda_{g}$
is the Pomeron intercept minus one and rises with $Q^{2}$
[10,13-15]. A set of formula to extract the gluon and the
structure function exponents was given in [15]. For our
calculations, We neglecting of the quark singlet part. So that,
the DGLAP equation for the gluon evolution in the NLO can be
written as:
\begin{equation}
Q^{2}\frac{{\partial}G}{{\partial}Q^{2}}=\frac{\alpha_{s}}{2\pi}{\int_{x}^{1}}
[P^{1}_{gg}(z)+\frac{\alpha_{s}}{2\pi}P^{2}_{gg}(z)]
G(\frac{x}{z},Q^{2})dz,
\end{equation}
where $P^{1}_{gg}(z)$ and $P^{2}_{gg}(z)$ are the LO and NLO
Altarelli- Parisi splitting kernels [12]. The running coupling
constant $\alpha_{s}(Q^{2})$ has the approximate analytical form
in NLO:
\begin{equation}
\frac{\alpha_{s}(Q^{2})}{2\pi}=\frac{2}{\beta_{0}\ln(\frac{Q^{2}}{\Lambda^{2}})}
[1-\frac{\beta_{1}\ln\ln(\frac{Q^{2}}{\Lambda^{2}})}{\beta_{0}^{2}\ln(\frac{Q^{2}}{\Lambda^{2}})}],
\end{equation}
where  $\beta_{0}=\frac{1}{3}(33-2N_{f})$ and
$\beta_{1}=102-\frac{38}{3}N_{f}$ are the one- loop (LO) and the
two- loop (NLO) correction to the QCD $\beta$- function, $N_{f}$
being the number of active quark flavours ($N_{f}=4$). Inserting
the splitting kernels $P^{1}_{gg}(x)$ and $P^{2}_{gg}(x)$ at the
small $x$ limit and carrying out the integration, we obtained [15]
an expression for $\lambda_{g}$ as follows:
\begin{eqnarray}
\ln\frac{\lambda_{g_{0}}}{\lambda_{g}-x^{\lambda_{g}}{\int_{t_{0}}^{t}}x^{-\lambda_{g}}(\frac{3\alpha}{\pi}-\frac{61\alpha^{2}}{9\pi^{2}})dt}\nonumber\\
={\int_{t_{0}}^{t}}(\frac{3\alpha}{\pi}-\frac{61\alpha^{2}}{9\pi^{2}})\frac{1-x^{\lambda_{g}}}{\lambda_{g}}dt.
\end{eqnarray}\\
Where
$\lambda_{g_{0}}$(=$\frac{{\partial}{\ln}G(x,t_{0})}{{\partial}{\ln}\frac{1}{x}}$)
is the exponent at the starting scale $t_{0}$  while $G(x,t_{0})$
is the input gluon distribution.\\

Also, to find an analytic solution for the singlet structure
function exponent, we note that the gluon term is dominate over
the scaling violation of $F_{2}$ at small- $x$. Neglecting the
quark, the DGLAP evolution equation for the singlet structure
function has the form:
\begin{equation}
\frac{dF^{S}_{2}}{dt}=\frac{\alpha_{s}}{2\pi}{\int_{x}^{1}}dz
(2N_{f}P_{qg}^{1}(z)+\frac{\alpha_{s}}{2\pi}P_{qg}^{2}(z))G(\frac{x}{z},Q^{2}).
\end{equation}
where $P^{1}_{qg}(z)$ and $P^{2}_{qg}(z)$ are the LO and NLO
Altarelli- Parisi splitting kernels [1,12]. After substitution and
rearrangement these equations, we obtained [15]:
\begin{eqnarray}
\lambda_{S}F_{2}(x,t)-\lambda_{S_{0}}F_{2}(x,t_{0})&=&\frac{0.555}{\pi}{\int_{t_{0}}^{t}}\alpha_{s}G(x,t)[(\frac{2\lambda_{g}}{3+\lambda_{g}}(1-x^{{3+\lambda_{g}}})+\frac{\lambda_{g}}{1+\lambda_{g}}(1-x^{{1+\lambda_{g}}})\nonumber\\
&&-\frac{2\lambda_{g}}{2+\lambda_{g}}(1-x^{{2+\lambda_{g}}}))+(2x^{{3+\lambda_{g}}}+x^{{1+\lambda_{g}}}-2x^{{2+\lambda_{g}}})]dt\nonumber\\
&&+\frac{1.852}{\pi^{2}}{\int_{t_{0}}^{t}}\alpha_{s}^{2}G(x,t)dt\hspace{4cm}
\end{eqnarray}
which defines the solution for $\lambda_{S}$. In this equation
$\lambda_{S_{0}}=\frac{{\partial}{\ln}F_{2}(x,t_{0})}{{\partial}{\ln}\frac{1}{x}}$
and  $F_{2}(x,t_{0})$ is input structure function at the starting
scale $t_{0}$.\\

Based on these results, we concentrate on the hard- Pomeron in our
calculations and present an approximation analytical solution for
the longitudinal structure function in the NLO corrections. We
test its validity comparing it with that of H1 data [8], Donnachie
$\&$ landshoff [16], MRST[17-19] and attempt to see how the
predictions for longitudinal structure
function are compared with the experimental data[8].\\

We specifically consider the next- to- leading- order (NLO)
corrections to the longitudinal structure function $F_{L}$,
projected from the hadronic tensor by combination of the metric
and the spacelike momentum transferred by the virtual photon
$(g_{\mu\nu}-q_{\mu}q_{\nu}/q^{2})$. In the next- to -leading
order the longitudinal structure function is proportional to
hadronic tensor as follows:
\begin{equation}
F_{L}(x,Q^{2})/x=\frac{8x^{2}}{Q^{2}}p_{\mu}p_{\nu}W_{\mu\nu}(x,Q^{2}),
\end{equation}
where $p^{\mu}(p^{\nu})$ is the hadron momentum and $W^{\mu\nu}$
is the hadronic tensor. In this relation we neglecting the hadron
mass.\\

The basic hypothesis is that the total cross section of a hadronic
process can be written as the sum of the contributions of each
parton type (quarks, antiquarks, and gluons) carrying a fraction
of the hadronic total momentum. In the case of deep- inelastic-
scattering it reads:
\begin{equation}
d\sigma_{H}(p)=\sum_{i}{\int}dyd\hat{\sigma}_{i}(yp)\Pi_{i}^{0}(y),
\end{equation}
where $d\hat{\sigma}_{i}$ is the cross section corresponding to
the parton $i$ and $\Pi_{i}^{0}(y)$ is the probability of finding
this parton in the hadron target with the momentum fraction $y$.
Now, taking into account the kinematical constrains one gets the
relation between the hadronic and the partonic structure
functions:
\begin{eqnarray}
f_{j}(x,Q^{2})=\sum_{i}{\int}_{x}^{1}\frac{dy}{y}\textsf{f}_{j}(\frac{x}{y},Q^{2})\Pi_{i}^{0}(y)=\sum_{i}\textsf{f}_{j}{\otimes}\Pi_{i}^{0}(y)\hspace{0.5cm},j=2,L,\\\nonumber
\end{eqnarray}
where $\textsf{f}_{j}(x,Q^{2})=F_{j}(x,Q^{2})/x$. Equation (10)
expresses the hadronic structure functions as the convolution of
the partonic structure function, which are calculable in
perturbation theory, and the probability of finding a parton in
the hadron which is a nonperturbative function. So, in
correspondence with Eq.(10) one can write Eq.(8) by follows:
\begin{eqnarray}
F_{L}/x&=&\frac{\alpha}{4\pi}[\textsf{f}_{L,q}^{(1)}{\otimes}(q_{S}^{0}+q^{0}_{NS})+\textsf{f}_{L,G}^{(1)}{\otimes}g^{0}]+(\frac{\alpha}{4\pi})^{2}[\textsf{f}_{L,q}^{NS(2)}{\otimes}(q_{S}^{0}+q^{0}_{NS})+\textsf{f}_{L,q}^{S(2)}{\otimes}q_{S}^{0}\nonumber\\
&&+\textsf{f}_{L,G}^{S(2)}{\otimes}g^{0}+O(\alpha^{3})],
\end{eqnarray}
where $q^{0}_{S}$ and $q^{0}_{NS}$ are the singlet and nonsinglet
quark distribution. $\textsf{f}_{L,q}^{(1)}$ and
$\textsf{f}_{L,G}^{(1)}$ are the LO partonic longitudinal
structure function corresponding to quarks and gluons,
respectively [21]. $\textsf{f}_{L,q}^{NS(2)}$ is the NLO quark
nonsinglet longitudinal structure function,
$\textsf{f}_{L,q}^{S(2)}$ is the NLO quark longitudinal structure
function which contributes only to the singlet, and
$\textsf{f}_{L,g}^{S(2)}$ is the NLO gluon longitudinal structure
function [20].\\

We present the expressions, after full agreement has been
achieved, in the form of kernels $K^{i}$, $i=NS, S,$ and $g$,
which give NLO- $F_{L}$ upon convolution with
$F_{2}=x\sum_{j=1}^{N_{f}}e_{j}^{2}(q+\overline{q})_{j}$,
$F_{2}^{S}=(\sum_{j=1}^{N_{f}}e_{j}^{2}/N_{f})x\sum_{j=1}^{N_{f}}(q+\overline{q})_{j}$,
and the gluon distribution $g$, respectively:
\begin{eqnarray}
F_{L}(x,Q^{2})&=&\int_{x}^{1}\frac{dy}{y}K^{NS}(\frac{x}{y},Q^{2})F_{2}(y,Q^{2})+\int_{x}^{1}\frac{dy}{y}K^{S}(\frac{x}{y},Q^{2})F_{2}^{S}(y,Q^{2})\nonumber\\
&&+\int_{x}^{1}\frac{dy}{y}K^{G}(\frac{x}{y},Q^{2})G(y,Q^{2}),
\end{eqnarray}
where $G(x,Q^{2})=xg(x,Q^{2})$ and $e_{j}$ are the quark charges
and $N_{f}$ the number of flavors [21]. The kernels have been
obtained with the modified minimal- subtraction ($\overline{MS}$)
scheme for the UV regularization combined with the DIS
prescription [20].\\

Based on  the Regge-like behavior for the gluon distribution and
singlet structure function, let us put Eqs.(1) and (2) in Eq.(12).
Thus Eq.(12) is reduced to:
\begin{eqnarray}
F_{L}(x,Q^{2})&=&\int_{x}^{1}\frac{dy}{y}[K^{NS}(\frac{x}{y},Q^{2})+K^{S}(\frac{x}{y},Q^{2})]A_{S}y^{-\lambda_{S}}\nonumber\\
&&+\int_{x}^{1}\frac{dy}{y}K^{G}(\frac{x}{y},Q^{2})A_{g}y^{-\lambda_{g}},\hspace{0.5cm}
\end{eqnarray}
Here the nonsinglet quark density is negligible at small $x$ and
the kernels $K^{i}$(i=nonsinglet, singlet and gluon) are defined
by:
\begin{eqnarray}
K^{NS}(\frac{x}{y},Q^{2})&=&\frac{\alpha_{s}}{4\pi}4C_{F}(x/y)^{2}+(\frac{\alpha_{s}}{4\pi})^2[4C_{F}(C_{A}-2C_{F})(x/y)^{2}[8(\frac{1}{2}ln(1-x/y)^2ln(+x/y)\nonumber\\
&&+ln(1-x/y)polylog(2,1-x/y)-polylog(3,1-x/y)+\zeta(3))+4polylog(3,x/y)\nonumber\\
&&+4polylog(3,-x/y)-4dilog(1+x/y)(ln(x/y)-2ln(1+x/y))\nonumber\\
&&-4ln(x/y)dilog(1-x/y)-2ln(x/y)^2ln(1-(x/y)^2)-8\zeta(3)+4ln(x/y)ln(1+x/y)^2\nonumber\\
&&+\frac{2}{5}(5-3(x/y)^2)ln(x/y)^2-\frac{23}{3}ln(1-x/y)-4\frac{2+10(x/y)^2+5(x/y)^3-3(x/y)^5}{5(x/y)^3}\nonumber\\
&&(+dilog(1+x/y)+ln(x/y)ln(1+x/y))+\frac{2}{3}Pi^2(ln(1-(x/y)^2)-\frac{5-3(x/y)^2}{5})\nonumber\\
&&+4\frac{6-3x/y+47(x/y)^2-9(x/y)^3}{15(x/y)^2}ln(x/y)-\frac{144+294x/y-1729(x/y)^2+216(x/y)^3}{90(x/y)^2}]\nonumber\\
&&+8C^{2}_{F}(x/y)^2[2dilog(1-x/y)+\frac{1}{2}ln(x/y)^2-\frac{Pi^2}{3}+\frac{41}{6}ln(x/y)-\frac{5}{3}ln(1-x/y)\nonumber\\
&&-\frac{96-535x/y}{36x/y}]-\frac{8}{3}C_{F}N_{f}(x/y)^2(ln(\frac{(x/y)^2}{1-x/y})-\frac{6-25x/y}{6x/y})],
\end{eqnarray}
\begin{eqnarray}
K^{S}(\frac{x}{y},Q^{2})&=&(\frac{\alpha_{s}}{4\pi})^2(\frac{16}{9}C_{F}N_{f}(3(1-2x/y-2(x/y)^2)(1-x/y)ln(1-x/y)\nonumber\\
&&+9(x/y)^2(+dilog(1-x/y)+ln(x/y)^2-Pi^2/6)+9x/y(1-2(x/y)^2)ln(x/y)\nonumber\\
&&-4(1-x/y)^3),
\end{eqnarray}
\begin{eqnarray}
K^{G}(\frac{x}{y},Q^{2})&=&\frac{\alpha_{s}}{4\pi}[8(x/y)^{2}(1-x/y)][\sum_{i=1}^{N_{f}}e_{i}^{2}]+(\frac{\alpha_{s}}{4\pi})^{2}[\sum_{i=1}^{N_{f}}e_{i}^{2}]16C_{A}(x/y)^2(+4dilog(1-x/y)\hspace{1cm}\nonumber\\
&&-2(1-x/y)ln(x/y)ln(1-x/y)+2(1+x/y)dilog(1+x/y)+3ln(x/y)^2\nonumber\\
&&+2(x/y-2)Pi^2/6+(1-x/y)ln(1-x/y)^2+2(1+x/y)ln(x/y)ln(1+x/y)\nonumber\\
&&+\frac{(24+192x/y-317(x/y)^2)}{24(x/y)}ln(x/y)+\frac{(1-3x/y-27(x/y)^2+29(x/y)^3)}{3(x/y)^2}ln(1-x/y)\nonumber\\
&&+\frac{(-8+24x/y+510(x/y)^2-517(x/y)^3)}{72(x/y)^2}-16C_{F}(x/y)^2(\frac{5+12(x/y)^2}{30}ln(x/y)^2\nonumber\\
&&-(1-x/y)ln(1-x/y)+\frac{-2+10(x/y)^3-12(x/y)^5)}{15(x/y)^3}(+dilog(1+x/y)\nonumber\\
&&+ln(x/y)ln(1+x/y))+2\frac{5-6(x/y)^2}{15}Pi^2/6+\frac{4-2x/y-27(x/y)^2-6(x/y)^3}{30(x/y)^2}ln(x/y)\nonumber\\
&&+\frac{(1-x/y)(-4-18x/y+105(x/y)^2)}{30(x/y)^2}).
\end{eqnarray}
For the SU(N) gauge group, we have $C_{A}=N$,
$C_{F}=(N^{2}-1)/2N$,
 $T_{F}=N_{f}T_{R}$, and $T_{R}=1/2$ where $C_{F}$ and $C_{A}$ are the color Cassimir operators. In our calculations, we use the Riemann $\zeta$ function
 and the well- known Nielsen generalized polytlogarithms, where the Nielsen$^{,}$s polylogarithm is defined by
 \begin{equation}
S_{n,p}(x)=\frac{(-1)^{n+p-1}}{(n-1)!p!}\int_{0}^{1}dt\frac{ln^{n-1}(t)ln^{p}(1-xt)}{t},
 \end{equation}
 In this equation, the values $n$ and $p$ are positive integers and $x$ is complex. Also the
 ordinary polylogarithm is given in terms of this as [22]:
\begin{equation}
Li_{n}(x)=S_{n-1,1}(x), n{\geq}2.
 \end{equation}
These equations are a set of formulas to extracted the
longitudinal structure function, using the gluon distribution
exponent and the structure function exponent determined in [15] at
small $x$ in the next- to- leading order of the perturbation
theory.\\

We computed the predictions for all detail of the longitudinal
structure function in the kinematic range where it has been
measured by $H1$ collaboration [8] and compared with DL model [16]
based on hard Pomeron exchange, also compared with computation
Moch, Vermaseren and Vogt [17-18] at the second order with input
data from MRST [19]. Our numerical predictions are presented as
functions of $x$ for the $Q^{2}=$12,15,20 and 25 $GeV^{2}$. The
average value $\Lambda$ in our calculations  is corresponding to
$292\hspace{0.1cm}MeV$. Results of these calculations are given in
Table.1. In Fig.1, the values of the NLO-longitudinal structure
functions are compare with the experimental $H1$ data[8]. The
curves represent the NLO QCD calculations
 $F_{L}$ based on a fit to the $1996-1997$ data. We compare our results with
 predictions of $F_{L}$ up to NLO in perturbative QCD [17-18] that
 the input densities is given by MRST parameterizations [19]. Also, we compare our results with the two pomeron fit as is
seen in Fig.1. These results indicate that the complete expression
for $F_{L}$ including the NLO corrections can provide solutions
for relevant tests of QCD. The data extend
 the knowledge of the longitudinal structure function into the
 region of low- $x$. This implies that the $x$ dependence of the
 longitudinal structure function at low $x$ is consistent with a
 power law, $F_{L}=A_{L}x^{-\lambda_{L}}$, for fixed $Q^{2}$.
 This behavior is associated with the exchange of an object known
 as the hard Pomeron. As can be seen in all figures, the increase of  our calculations for the
 longitudinal structure functions $F_{L}(x,Q^{2})$ towards low
 $x$ are consistent with the NLO QCD calculations.\\

 Based on Regge- like behavior of the longitudinal structure
 function, we calculate exponent $\lambda_{L}$ and compare our
 results with the experimental results from $H1$ Collaboration [8]
 that given as the derivative of the longitudinal structure
 function with respect to $\ln\frac{1}{x}$, shows by
\begin{equation}
\lambda_{L}=\frac{{\partial}{\ln}F_{L}(x,Q^{2})}{{\partial}{\ln}\frac{1}{x}}|_{Q^{2}=cte}.
 \end{equation}
The result of calculation is shown in Fig.2. In this figure, we
show $\lambda_{L}$ calculated as a function of $Q^{2}$. Our
results show that $\lambda_{L}$ is independent of $x$ but has a
negative slope with respect to $t[=\ln\frac{Q^{2}}{\Lambda^{2}}]$.
The result for $A_{L}(Q^{2})$ is presented in Fig.3. The
coefficients $A_{L}(Q^{2})$ are  dependence  of $t$ and increase
linearly. Having concluded that the data for $F_{L}$ require a
hard Pomeron component, it is necessary to test this with our
results.\\

In conclusion, in this paper we have obtained an analytic solution
for the longitudinal structure function in the next- to- leading
order at low $x$. We found that the Regge theory can be used to
constrain the hard Pomeron exchange to the longitudinal structure
function behavior. To confirm the method
 and results, the calculated values are compared with the $H1$ data on the longitudinal
 structure function, at small $x$ and QCD fits. These results implies that the NLO contributions improve substantially
 the agreement with the QCD fit. Thus implying that Regge theory and perturbative evolution may be
 made compatible at small $x$. Thus, this behavior at low $x$ is consistent
with a dependence $F_{L}(x,Q^{2})=A_{L}x^{-\lambda_{L}}$
throughout that region. The longitudinal structure function
increase as usual, as $x$ decreases. The form of the obtained
distribution function for the longitudinal structure function is
similar to the predicted from the proton paramerterization, and
this is in agreement with the increase observed by the $H1$
experiments. Also, in this paper we have obtained $\lambda_{L}$ in
the next- to leading order at low $x$. our results show that the
derivatives
$\frac{{\partial}{\ln}F_{L}(x,Q^{2})}{{\partial}{\ln}\frac{1}{x}}=\lambda_{L}(x,Q^{2})$
is independent of $x$. At low $x$, the exponent $\lambda_{L}$ has
a negative slope with respect to $t$ and the coefficient $A_{L}$
is observed to rise linearly with $t$. This behavior of the
longitudinal structure function at low $x$ is consistent with a
power- law behavior.\\

\newpage
\textbf{References}\\
\hspace{2cm}1. G.G.Callan and D.Gross, Phys.Lett.B\textbf{22}, 156(1969);\\
\hspace{2cm}2. R.G.Roberts, The structure of the proton, (Cambridge University Press 1990)Cambridge.\\
\hspace{2cm}3. A.V.Kotikov, JETP Lett.\textbf{59}, 1(1994); Phys.Lett.B\textbf{338}, 349(1994).\\
\hspace{2cm}4. Yu.L.Dokshitzer, D.V.Shirkov, Z.Phys.C\textbf{67}, 449(1995).\\
\hspace{2cm}5. W.K.Wong, Phys.Rev.D\textbf{54}, 1094(1996).\\
\hspace{2cm}6. S.Aid et.al, $H1$ collab. phys.Lett. {\bf B393}, 452-464 (1997).\\
\hspace{2cm}7. R.S.Thorne, phys.Lett. {\bf B418}, 371(1998).\\
\hspace{2cm}8. C.Adloff et.al, $H{1}$ Collab., Eur.Phys.J.C\textbf{21}, 33(2001).\\
\hspace{2cm}9. N.Gogitidze et.al, $H{1}$ Collab., J.Phys.G\textbf{28}, 751(2002).\\
10. A.V.Kotikov and G.Parente, JHEP \textbf{85}, 17(1997);
Mod.Phys.Lett.A\textbf{12}, 963(1997).\\
11. C.Adloff,$H1$ collab. phys.Lett. {\bf B393}, 452(1997).\\
12. Yu.L.Dokshitzer, Sov.Phys.JETP {\textbf{46}}, 641(1977);
G.Altarelli and G.Parisi, Nucl.Phys.B \textbf{126}, 298(1977);
V.N.Gribov and L.N.Lipatov,
Sov.J.Nucl.Phys. \textbf{15}, 438(1972).\\
13. M.Gluk, E.Reya and A.Vogt, Euro.J.Phys.C\textbf{5}, 461(1998).\\
14. A.D.Martin, W.S.Striling and R.G.Roberts, Euro.J.Phys.C {\bf 23}, 73(2002).\\
15. G.R.Boroun and B.Rezaie, Phys.Atom.Nucl.vol.71, No.6, 1076(2008) \\
16. A. Donnachie and P.V.Landshoff, Phys.Lett.B\textbf{533},
277(2002); Phys.Lett.B\textbf{550}, 160(2002);\\ J.R.Cudell, A.
Donnachie and P.V.Landshoff, Phys.Lett.B\textbf{448}, 281(1999);\\
P.V.Landshoff, hep-ph/0203084.\\
17. A.Vogt, S.Moch, J.A.M.Vermaseren, Nucl.Phys.B \textbf{691},
129(2004).\\
18. S.Moch, J.A.M.Vermaseren, A.vogt, Phys.Lett.B \textbf{606},
123(2005).\\
19. A.D.Martin, R.G.Roberts, W.J.Stirling,R.Thorne, Phys.Lett.B \textbf{531}, 216(2001).\\
20. J.L.Miramontes, J.sanchez Guillen and E.Zas, Phys.Rev.D \textbf{35}, 863(1987).\\
21. D.I.Kazakov, et.al., Phys.Rev.Lett. \textbf{65}, 1535(1990).\\
22. A.Devoto, et.al., Phys.Rev.D \textbf{30}, 541(1984).\\
23. A.M.Cooper-Sarkar and R.C.E.Devenish, Acta.Phys.Polon.B34,
 2911(2003).\\
24. R.K.Ellis , W.J.Stirling and B.R.Webber, QCD and Collider
Physics(Cambridge University
Press)1996.\\


\begin{widetext}
\setlength{\tabcolsep}{4pt}
\begin{table}[h]
\centering \caption{The longitudinal structure function terms
based on Regge-like behavior.}\label{table:table1}
\begin{minipage}{\linewidth}
\renewcommand{\thefootnote}{\thempfootnote}
\centering
\begin{tabular}{|l|c||c|c|c|c|l|}
\hline\noalign{\smallskip} $Q^{2}(GeV^{2})$ &  $ x $ &  $ LO-gluon
 $& $ NLO-gluon$ & $NLO-singlet$&$ LO-nonsinglet$&$NLO-nonsiglet$\\
\hline\noalign{\smallskip}
12 & 0.000161 & 0.429 & -0.029 & -0.070 & 0.072 & 0.025 \\
12 & 0.000197 & 0.406 & -0.026 & -0.066 & 0.067 & 0.023 \\
12 & 0.000320 & 0.358 & -0.019 & -0.057 & 0.059 & 0.020 \\
15 & 0.000201 & 0.431 & -0.025 & -0.066 & 0.072 & 0.024 \\
15 & 0.000246 & 0.407 & -0.023 & -0.062 & 0.068 & 0.022 \\
15 & 0.000320 & 0.383 & -0.019 & -0.057 & 0.063 & 0.020 \\
20 & 0.000268 & 0.384 & -0.018 & -0.062 & 0.074 & 0.023 \\
20 & 0.000328 & 0.367 & -0.016 & -0.057 & 0.068 & 0.021 \\
20 & 0.000500 & 0.333 & -0.012 & -0.050 & 0.060 & 0.019 \\
25 & 0.000335 & 0.383 & -0.015 & -0.056 & 0.072 & 0.022 \\
25 & 0.000410 & 0.364 & -0.013 & -0.052 & 0.068 & 0.020 \\
25 & 0.000500 & 0.348 & -0.010 & -0.049 & 0.064 & 0.019 \\
\hline\noalign{\smallskip}
\end{tabular}
\end{minipage}
\end{table}
\end{widetext}
\begin{figure}
\includegraphics[width=1.1\textwidth]{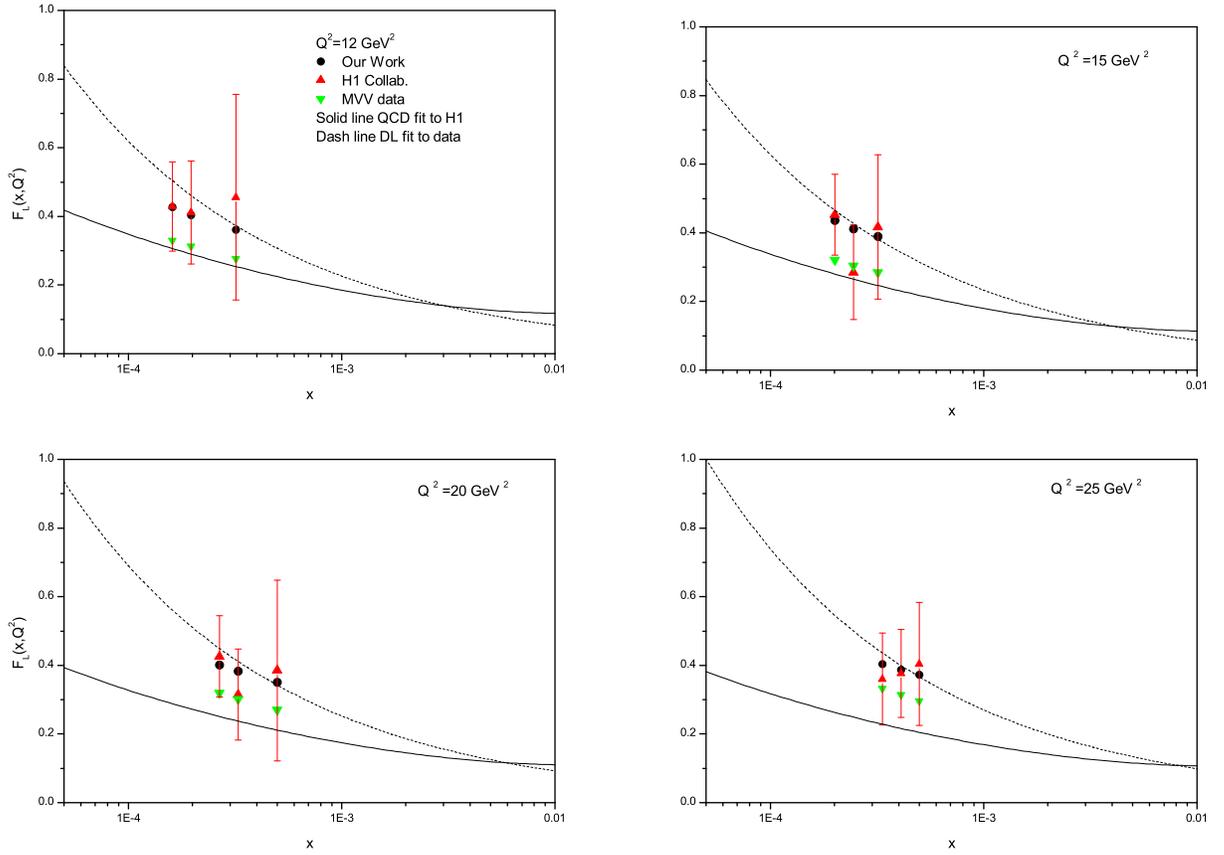}
\caption{H1 data [8](up triangle) for the longitudinal structure
function at $Q^{2}$=12,15,20 and 25\hspace{0.1cm}$GeV^{2}$ values,
with our NLO data calculations. The error on the  H1
 data is the total uncertainty of the determination of
 $F_{L}$ representing the statistical, the systematic and the model errors added in quadrature.
Down triangle data are the MVV prediction [18]. The solid line is
the NLO QCD fit to the H1 data for $y<0.35$ and
  $Q^{2}{\geq}3.5\hspace{0.1cm}GeV^{2}$. The dash line is the DL [16] fit to $F_{L}$.}
  \label{Fig1}
\end{figure}
\begin{figure}
\includegraphics[width=0.5\textwidth]{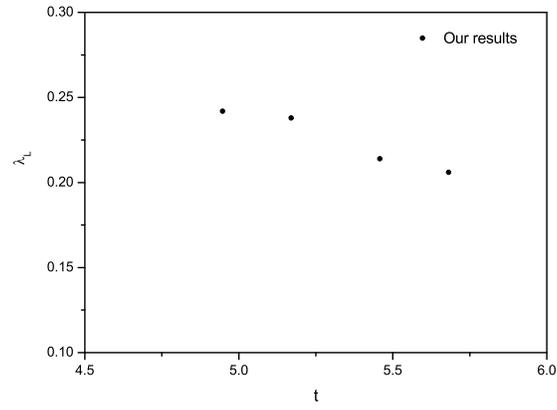}
\caption{Calculation of the exponent $\lambda_{L}$ from fits of
the form $F_{L}(x,Q^{2})=A_{L}x^{-\lambda_{L}}$ to our
longitudinal structure function data.} \label{Fig2}
\end{figure}
\begin{figure}
\includegraphics[width=0.5\textwidth]{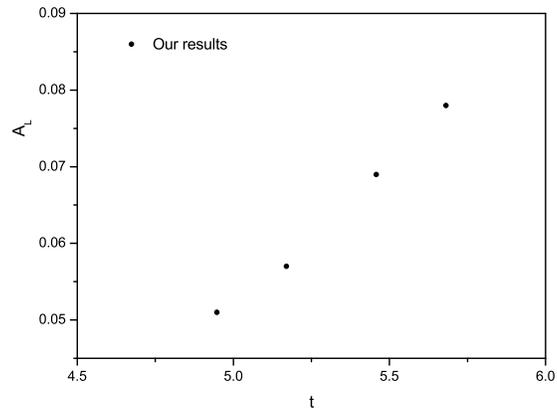}
\caption{Calculation of the coefficient $A_{L}$ from fits of the
form $F_{L}(x,Q^{2})=A_{L}x^{-\lambda_{L}}$ to our longitudinal
structure function data.} \label{Fig3}
\end{figure}
\end{document}